# Deep learning polarization distributions in ferroelectrics from STEM data: with and without atom finding

Christopher T. Nelson,[1,a] Ayana Ghosh,[1,2,b] Mark Oxley,[1] Xiaohang Zhang,[3] Maxim Ziatdinov,[1,2] Ichiro Takeuchi,[3] and Sergei V. Kalinin[1,c]

[1] Center for Nanophase Materials Sciences, Oak Ridge National Laboratory, Oak Ridge, TN 37831

[2] Computational Sciences and Engineering Division, Oak Ridge National Laboratory, Oak Ridge, TN 37831

[3] Department of Materials Science and Engineering, University of Maryland, College Park, MD 20742

Over the last decade, scanning transmission electron microscopy (STEM) has emerged as a powerful tool for probing atomic structures of complex materials with picometer precision, opening the pathway toward exploring ferroelectric, ferroelastic, and chemical phenomena on the atomic-scale. Analyses to date extracting a polarization signal from lattice coupled distortions in STEM imaging rely on discovery of atomic positions from intensity maxima/minima and subsequent calculation of polarization and other order parameter fields from the atomic displacements. Here, we explore the feasibility of polarization mapping directly from the analysis of STEM images using deep convolutional neural networks (DCNNs). In this approach, the DCNN is trained on the labeled part of the image (i.e., for human labelling), and the trained network is subsequently applied to other images. We explore the effects of the choice of the descriptors (centered on atomic columns and grid-based), the effects of observational bias, and whether the network trained on one composition can be applied to a different one. This analysis demonstrates the tremendous potential of the DCNN for the analysis of high-resolution STEM imaging and spectral data and highlights the associated limitations.

---

[a] Equal contributors
[b] Equal contributors
[c] Corresponding author, sergei2@ornl.gov

The functionality of ferroelectric materials is inseparably linked to the static distributions and dynamic behaviors of the polarization.[1-5] The discontinuity of polarization is associated with the emergence of bound charge, resulting in strong coupling between the polarization and electrochemical,[6-13] semiconductive,[14-17] and transport phenomena.[18-24] Compared to ferromagnets, ferroelectrics have extremely short correlation lengths and domain wall widths, on the order of several unit cells. This results in an extreme sensitivity of the polarization dynamics on the atomic structure. For example, since the early work of Miller and Weinreich[25] and Burtsev and Chervonobrodov[26-28] it has been realized that domain wall motion proceeds via the generation of kinks in the domain walls. This further results in strong interactions between topological defects in ferroelectrics and charged impurities, giving rise to unique functionalities of ferroelectric relaxors.[29-31]

These considerations have stimulated extensive efforts toward exploring ferroelectric materials on the atomic level via (scanning) transmission electron microscopy, (S)TEM. The feasibility of visualizing polarization fields by TEM was first demonstrated in the late 1990s by Pan.[32] A decade later, work by Jia demonstrated the potential of TEM for mapping polarization behavior at the level of individual structural[33] and topological[34, 35] defects. At about the same time, groups at Oak Ridge National Laboratory[36-38] and the University of Michigan[39] demonstrated STEM imaging of polarization in ferroelectrics, igniting rapid growth in this field. In these studies, STEM data is used to directly position the centroids of atomic columns and then the unit-cell-scale dipoles are calculated from the product of the displacements with associated Born or Bader charges.[40] Multiple observations of polarization distribution on topological defects,[41-43] interfaces,[44] modulated structures,[45] and extended defects[33, 46] have been reported.

These studies have not only offered visualization of the polarization fields but have also allowed quantitative insights into the physics of ferroelectric materials. In the mesoscopic Ginzburg-Landau models, the structure of polarization distributions in the vicinity of domain walls or interfaces is intrinsically linked to the structure of the free energy functional, its gradient or flexoelectric terms, and the boundary conditions.[47-49] Correspondingly, quantitative analysis of STEM data can provide insight regarding the corresponding mechanisms.[43, 50] Recently, this analysis has been extended toward the Bayesian analysis of domain wall structures, allowing incorporation of past knowledge of materials physics into the model and quantifying the requirements to microscopic systems required to identify specific aspects of physical behaviors.[51]

These analyses necessitate understanding of the veracity of the polarization analysis from STEM images and further necessitate the development of image analysis tools that allow rapid transformation of the STEM images into polarization fields, both as a first step toward physics-based analyses and as a necessary step toward automated experimentation with image-based feedback. Here, we explore the applications of deep convolutional neural networks (DCNNs) for reconstruction and segmentation of STEM images of ferroelectric materials and explore some of the potential sources of observational biases in this analysis.

As a model system we explore a thin film of the Sm-doped ferroelectric $BiFeO_3$ (BFO) epitaxially grown on a $SrTiO_3$ (STO) substrate as a combinatorial library with Sm concentration varying from 0 to 20%. Several $Sm_xBi_{1-x}FeO_3$ STEM samples with different substitution concentration x are obtained from one composition spread[51, 52] spanning x = 0% (pure $BiFeO_3$ (BFO)) to 20% ($Bi_{0.8}Sm_{0.2}FeO_3$). For BFO the ferroelectric polarization strongly couples with the lattice, notably the heavy cation Bi and Fe sublattices which are readily imaged by atomic-resolution STEM, and this cation non-centrosymmetry is used as a proxy for the ferroelectric polarization vector. STEM images are collected using a high-angle annular dark field (HAADF) detector, which for zone-axis projected crystalline materials produce intuitive bright-atom contrasts images such as that shown in Figure 1a for $[100]_{psuedocubic}$ BFO. The growth parameters, sample preparation, and imaging details are the same as in our previous publications.[51, 52] The data set for the composition series is publicly available at DOI 10.5281/zenodo.4555978.

The spatial distribution of lattice structures and symmetry breaking distortions can be derived from the real-space positions of the atoms, relying on parameterizations of the atomic columns that are typically fitted as Gaussians. This process is illustrated in Figure 1 for mapping the distribution of polarization in pure rhombohedral BFO that manifest in a phase offset between the local Bi A-site and Fe B-site sublattices. Figure 1 (b) depicts a local neighborhood of Gaussian fits corresponding to the inset HAADF-STEM image. This polar displacement vector, *P*, is defined as the difference between the central Fe (red) position and the average of the four neighbor Bi atom positions (blue) or vice-versa for a Bi-centered cell. The colorized vector distribution for the HAADF-STEM image is shown in Figure 1 (c), illustrating the polydomain polarization distribution, which is dominated by a 109°-type domain wall bisecting the image. An upper bound of the positional error of this parameterization can be made by measuring the total variation from the ideal uniform lattice spacings. Heat maps of the variation from mean values are shown for the Bi A-site and Fe B-site sublattices in Figure 1 (d) for a distance of 1 Å with RMS values of 9.7 pm and 14.6 pm, respectively, illustrating the greater uncertainty of fitting the dimmer Fe positions. This is also apparent from uncertainty estimates from the Gaussian-fit optimization function, as shown in the histograms of Figure 1 (e). Some measurements can be made on the higher precision A-site sublattice alone, such as lattice spacings/strain, but the polar displacement requires both sublattices and thus, the Fe site is the most significant error contribution. The spatial distribution of $P$ error estimates from constituent atomic fitting is shown in Figure 1 (f) showing that the uncertainty systematically increases in some regions such as the STO substrate at the top and closer to the free surface (bottom of image, especially at right).

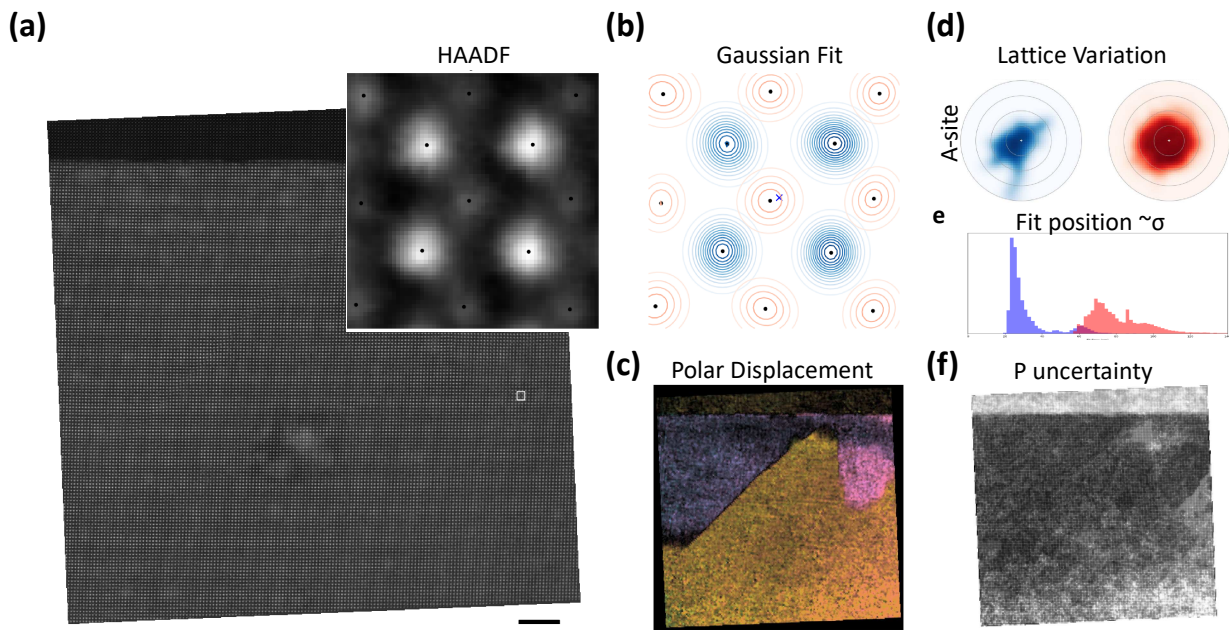

**Figure 1**. Polarization mapping via atomic position parametrization. a) HAADF-STEM image of a [100]psuedocubic BFO film on STO substrate (top). Fe-centered perovskite unit-cell is shown inset. (b) Gaussian fits corresponding to inset region. Polar displacement, $P$, defined as the vector between central Fe position and average Bi position (blue cross). (c) $P$ distribution map corresponding to HAADF-STEM image in (a), principal features are 109° (blue/gold) and 180° (gold/pink) domain walls. Color legend maximum radius corresponds to 50pm. (d) Heat map of variation from uniformity for Bi A-site (blue) and Fe B-site (red) sublattices, each band corresponds to 25pm. RMS values are 9.7 pm and 14.6 pm, respectively. (e) Histogram of uncertainty estimates from fitting for Bi A-sites (blue) and Fe B-sites (red). (f) Distribution of $P$ uncertainty. Scalebar is 5 nm.

The process of polarization field mapping by this approach is computationally intensive, requiring identifying all the atoms in the system, a fitting refinement of their position, and mapping neighbor relationships. In practice manual input is often necessary too in order to curate, threshold, filter/smooth, set parameter fitting bounds, remove lattice defects, etc. Furthermore, as with any point estimate it is also associated with relatively high noise. Similarly, the use of the ad-hoc Gaussian fitting to position the atomic column center as opposed to deconvolution using the correct beam profile leads to systematic fitting errors. Finally, measurement artifacts associated with zone-axis mis-tilt can also manifest as sublattice phase offsets, leading to systematic errors of this measurement that are independent of polarization values[53, 54]. In practice this leads to observed polarization values of opposite domains mirrored at a domain wall to exhibit unequal magnitudes, or the appearance of non-centrosymmetry in centrosymmetric materials.

We explore the applications of supervised DCNNs for the extraction of polarization and other structural descriptors from STEM image data with and without atom finding. All details of this framework can also be found in the accompanying Jupyter notebooks. As a first step, we establish whether DCNN analysis can substitute for classical featurization of the STEM images if atomic positions are predetermined, e.g., using deep learning atom finding algorithms.[55, 56]

Here, we implement the PyTorch DCNN models with three convolution blocks; the first one contains five 2D convolution layers with 32 filters each; the second has two 2D convolution layers with 64 filters each; and the third has one convolution block with two 2D convolution layers having 128 filters each. The leaky rectified linear unit (LReLU) is considered as the activation function in all these blocks. A 2D max pooling layer for dimensionality reduction is also added at the end of the second convolution block. A dropout for preventing overfitting and a batch normalization layer for training networks in mini batches are added toward the very end of the network architecture. The feature set is the sub-images (80*80), whereas the target vector is the unit-cell descriptors such as unit-cell parameters, volume, and polarization vector components.

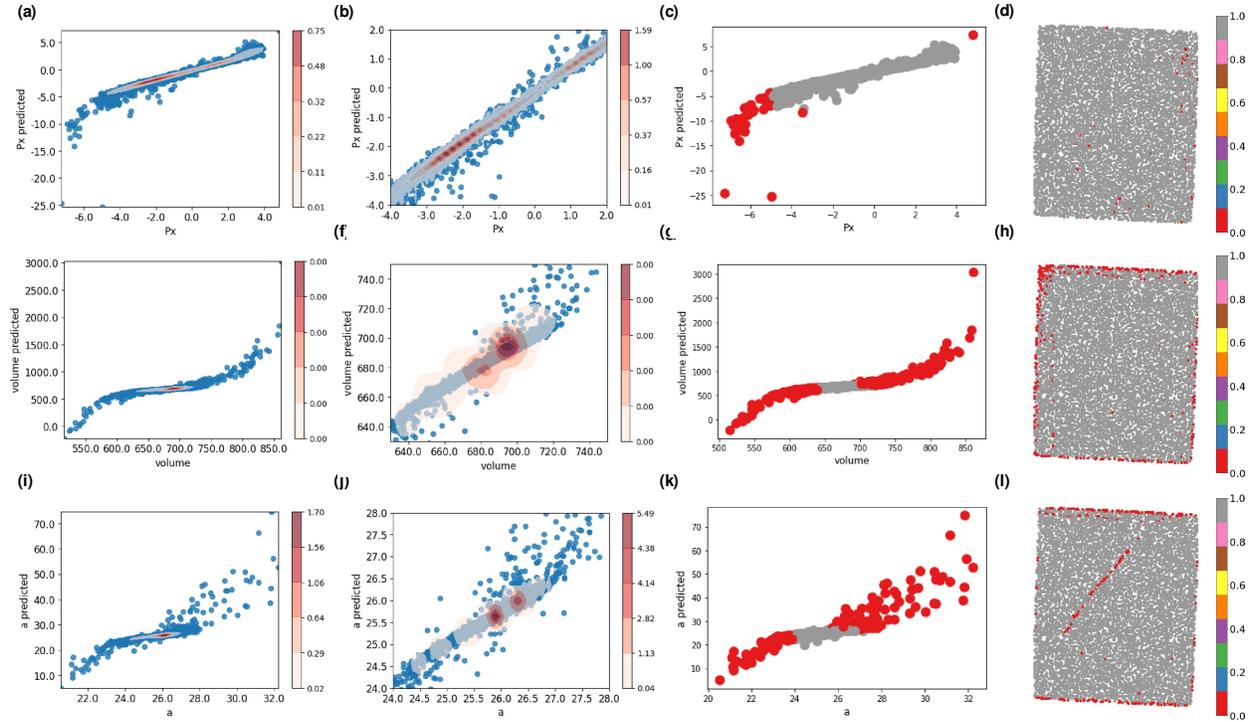

**Figure 2**. DCNN predicted vs. measured polarization, $\boldsymbol{P}_x$, component in (a) full and (b) zoomed-in scale. DCNN predicted vs. measured projected unit-cell volume in (e) full and (f) zoomed-in scale. DCNN predicted vs. measured unit-cell size in (i) full and (j) zoomed-in scale. The anomalous regions along with their respective spatial locations (both marked in red) associated with the predictions are also shown in (c-d), (g-h) and (k,l) for measured $\boldsymbol{P}_x$, measured unit-cell volume and measured unit-cell size, respectively.

Figure 2 shows the comparisons of the of the DCNN predictions and the ground truth data. The individual points and kernel density estimates for the distribution are shown as a way to visualize both the average behaviors and outliers. The observed dynamics are rather remarkable. For the majority of the locations, the DCNN-predicted parameters tend to have narrower distributions then the original (measured) values. This behavior is expected since DCNNs tend to smooth the data. However, for extreme values of the parameters the DCNN predictions start to deviate strongly, leading to unphysical predicted values.

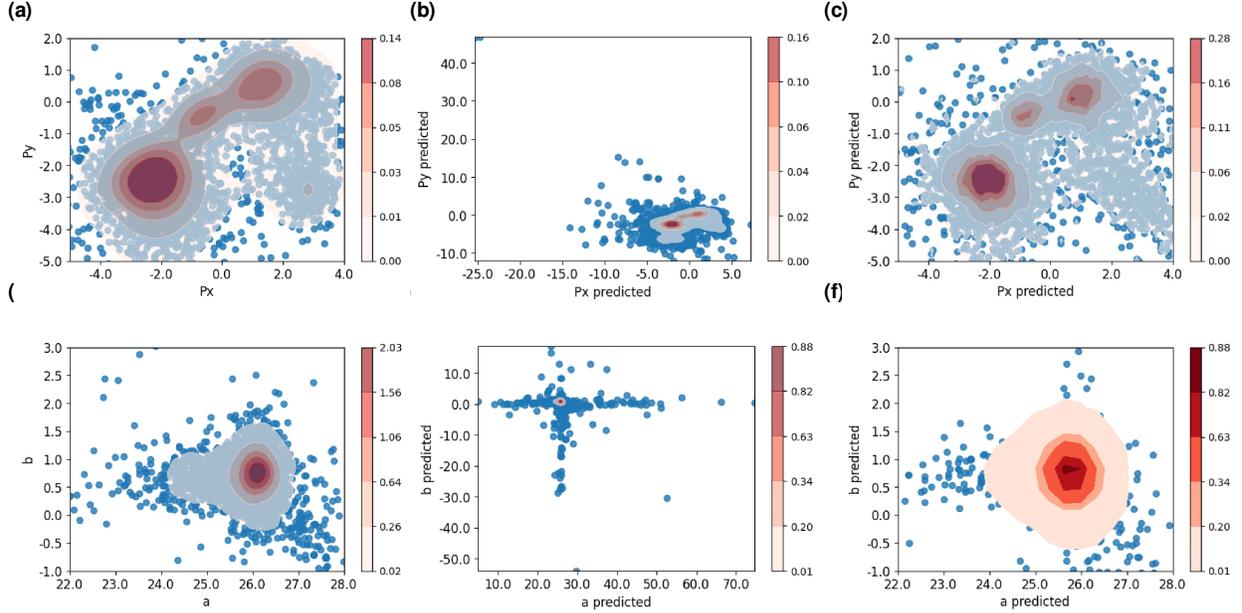

**Figure** 3. Joint distribution of $P_x$ vs. $P_y$ components (a) measured, (b) DCNN-predicted, and (c) zoomed-in of (b). Joint distribution of lattice parameters, a and b, (d) measured, (e) DCNN-predicted, and (f) zoom-in in (e).

Further shown in Figure 3 are the joint distributions of the measured and DCNN-predicted polarization components. The distribution is clearly multimodal, reflecting the ferroic variants present in the system. Remarkably, the observed maxima are asymmetric, suggesting that the polarization values extracted from the STEM images contain systematic errors, as from mistilt.[53,54] The corresponding distributions for the *a* and *b* lattice parameters are shown in Figure 3 (b) where the maxima corresponding to the film and substrate are clearly seen.

While the analyses in Figures 2 and 3 show reduced noise levels compared to classical analyses, they only offer a partial advantage compared to the classical approach since both are based on identification of atomic position. Here, we further explore whether the DCNN approach can be used for mapping polarization fields in the raw STEM images without using atom finding. We note that this is expected to be feasible given the DCNNs are invariant to translations in the image plane.

To explore this, we configured a 'sliding window' approach to generate sub-images that are not necessarily centered around specific atoms. For a predefined window size, parts of the STEM images lying inside the window are first considered. These form the feature set for the DCNN training, i.e., local descriptor. To create the target set, i.e., the corresponding polarization or unit-cell volume value, we adopt the following approach. An upper bound of the cation-cation average interatomic distance along with a minimum distance to the centers of the identified unit-cells are provided. The upper bound signifies the distance at which the contribution of a unit-cell to polarization becomes zero. Next, a KDTree is employed to query all the closest neighbors for a given list of coordinates to select only those falling within the specified maximum distance. The resulting $(x,y)$ coordinates, corresponding polarization values along the $x, y$ directions, and sub-images are utilized for training.

Each stack of sub-images and the associated physical values for different concentrations of Sm are used to build the various networks. A set of 26 networks are constructed for all 13 stacks of sub-images (generated from individual input images) with a window size of 80 pixels, centered (C) and not centered (NC) around atoms utilizing the same architecture as above.

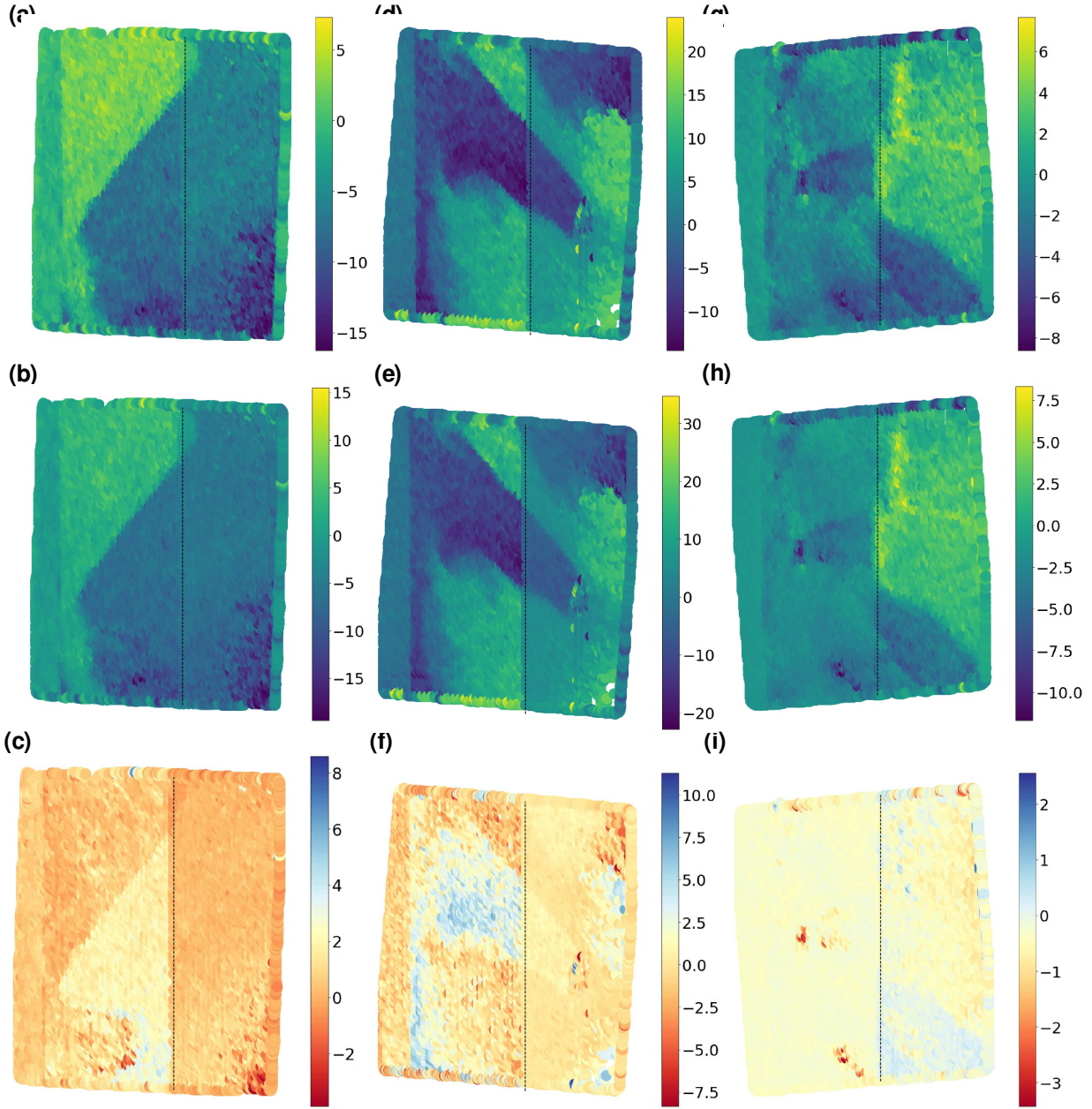

**Figure 4.** Polarization maps (using NC) representing the ground truth (top row), prediction (middle row), and differences between them (bottom row) are shown for 0% (a-c), 7% (d-f) and 10% (g-i) Sm-doped BFO, respectively. Predictions obtained using three different networks as trained on 2/3 of the full stack of sub-images (for every concentration) and tested on the rest. Vertical line in plots refer to the train-test splits.

Polarization maps are generated by plotting the measured and predicted polarization values, as shown in Figure 4. In particular, maps in (a-c), (d-f), and (g-i) represent the true polarization values, predicted ones by networks trained on the same Sm concentrations, and point-by-point difference between them showing uncertainties in predictions for 0%, 7%, and 10% Sm concentrations, respectively. Note that while 0% Sm corresponds to the pure rhombohedral ferroelectric BiFeO3, the 7-10% doping corresponds to the monoclinic phases at the morphotropic boundary and 20% corresponds to orthorhombic non-ferroelectric phase. In all cases, the uncertainty is relatively low, assuring reasonable performance of these networks. How extendable these predictions are (as discussed later), meaning if trained on one and applied to another can lead to similar accuracies, is also extremely important to further show robustness of such networks.

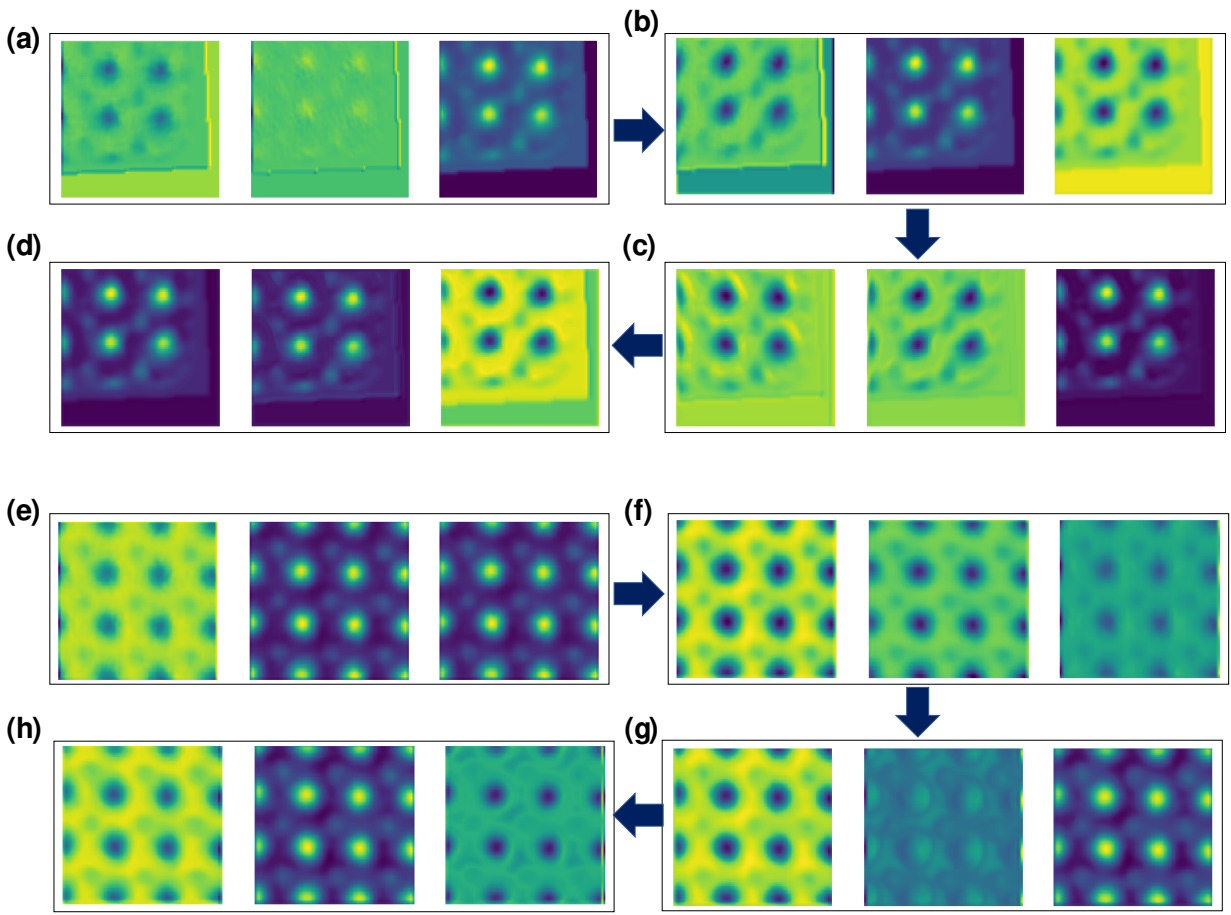

**Figure 5.** Representative feature maps for three different filters of the first four convolution layers of DCNNs trained on sub-images of 0% Sm-doped BFO are shown. The first block (a-d) corresponds to the network trained with C while the bottom block (e-h) refers to another model trained with NC.

To gain insight into the DCNN operations, we constructed feature maps for individual trained DCNNs illustrating how the input is transformed passing through the convolution layers. Once an input image is passed through a specific block, layer, and filter, the immediate activations are recorded, which are plotted to visualize the corresponding encoded features. For each layer, there are multiple (32 or 64 or 128) filters yielding individual feature maps. For example, for one convolution layer with 32 filters, a sum of 32 feature maps can be plotted corresponding to each filter for that specific layer. Figure 5 shows selected feature maps for four convolution layers of the first block of the networks. The DCNNs that are trained on a stack of sub-images that are both NC (a-d) and C (e-h) around atoms are utilized for constructing these representative maps. From these feature maps, it is evident that atoms in both lattices become more prominent in each filter as we progress from one layer to the next one.

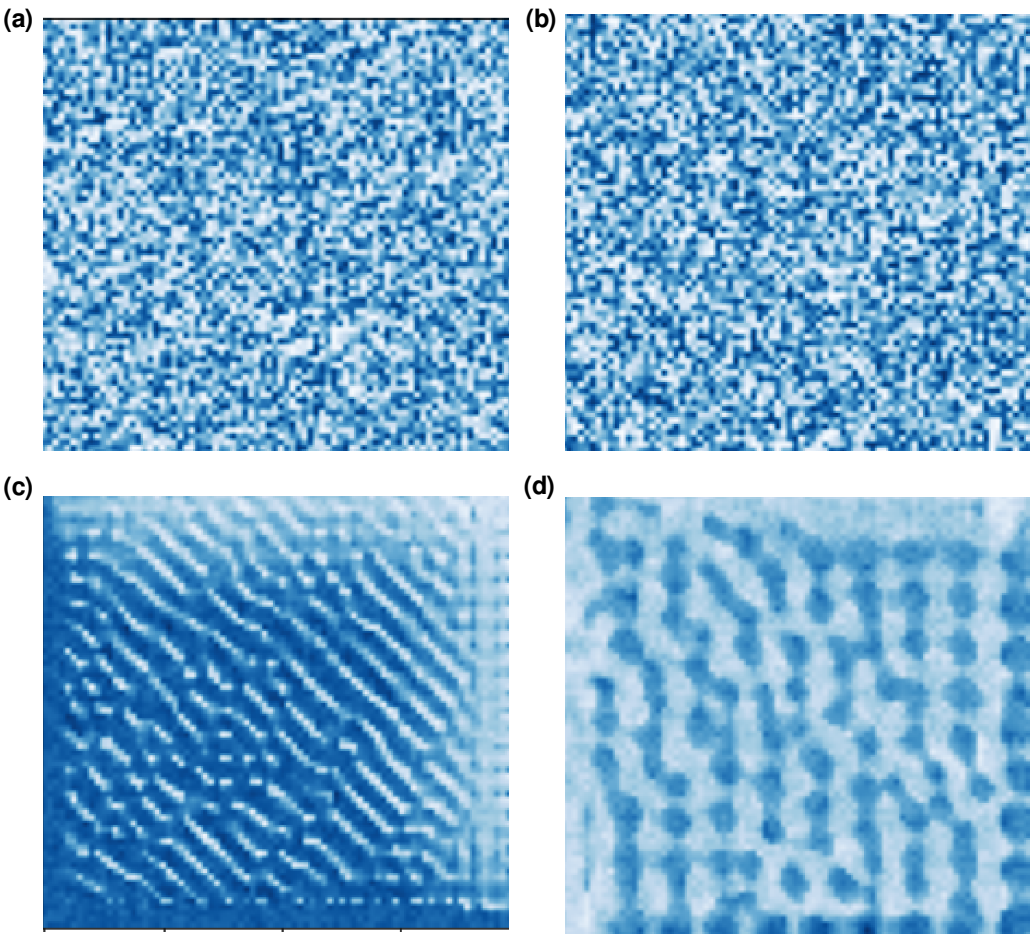

**Figure 6**: Random image (a) and activations of three filters of three consecutive CNN layers of a DCNN trained with NC are visualized in (b-d).

In addition to feature maps, we also visualized CNN filters present in different blocks (similar to the celebrated DeepDream[57] approach). These visualizations primarily display the patterns each filter maximally respond to. Any random image (could be one from one of the sub-image stacks) is considered as input. A loss function maximizing the value of the CNN filter is used to iteratively perform gradient ascent in the input space such that the algorithms find input values where the filter is activated the most. Figure 6 has a few representative visualizations of how the first three layers (b-d) of the first convolution block are activated as a random image (a) is selected as an input to this specific network.

The activations in the last kernel for three consecutive filters are shown in Figure 6. This analysis not only helps to understand the network architecture in greater detail but also shows how layers located deeper in the network facilitate in visualizing more training data-specific features. In the specific example in Figure 6, the network follows the same trend where visualization of the third layer (d) displays more patterns as compared to that in (b) or (c). Feature maps for all the filters for all the convolution layers present in each convolution block, as well as examples of CNN filter visualization of different blocks, can be found in the accompanying Jupyter Notebooks.

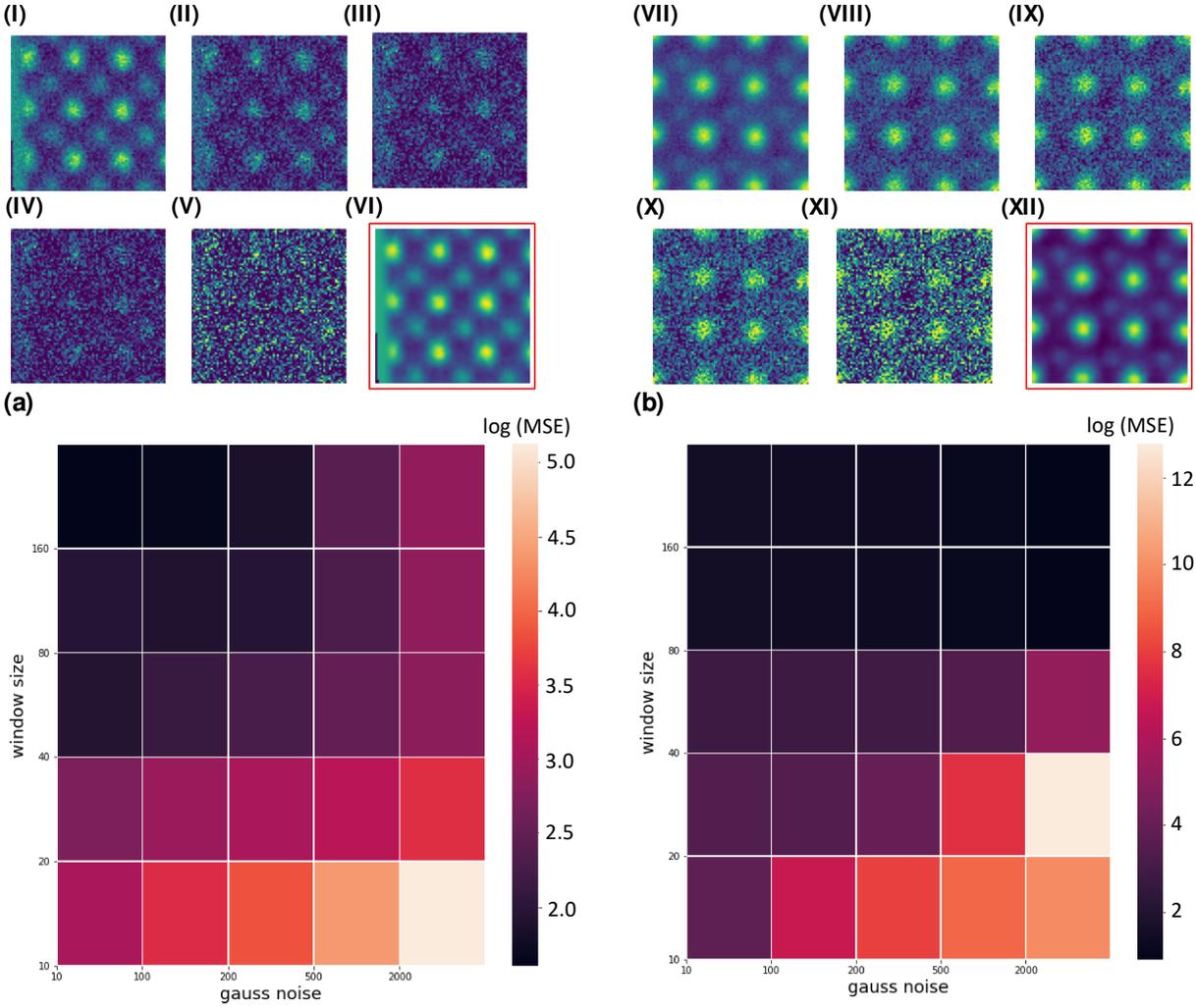

**Figure 7.** Noise sensitivity/performance of DCNNs is shown for different window sizes (10, 20, 40, 80, and 160 pixels) when various levels of gauss noise (10, 100, 200, 500, 2000 in arb. units) are added to the dataset. Selective sub-images for window size of 80 with various gauss-noises added are shown in (I-V) and (VII-XI) for NC and C. Reference sub-images are represented by VI and XII corresponding to sub-images generated with no added noise with the same window size. DCNNs trained with a stack of sub-images with no noise added are utilized to predict on this group of noisy sub-images. MSE values for each window size-noise combination are plotted as heatmaps in (a) and (b) for one image for both NC and C, respectively.

To test the extendibility and applicability of DCNNs trained with the best available dataset comprising NC and C sub-images, these networks are utilized to also predict polarization values for a set of noisy images. A set of noisy images for both NC and C sub-images are generated by adding five different magnitudes of gauss noise (GN) such as (10, 100, 200, 500, 2000 in arb. units) for varied window sizes of 10, 20, 40, 80, and 160. Here, each given value such as 10, 100 etc. is multiplied by $10^{-4}$ and passed via scikit random noise function to add GN to the images. An example of how the noisy images appear is shown in Figure 7, where (I-V) and (VII-XI) are the noisy NC and C sub-images (frame #0), respectively, with a window size of 80 with GNs added individually from the set of noises. Reference images of the same window size and sub-images with no added noise are shown in VI and XII. The DCNNs for each window size as trained on a dataset without any additional noise introduced to the sub-images are then considered as pre-trained models to evaluate the mean square error (MSE) values of predicted polarization corresponding to noisy sub-images and real polarization values. The error matrices (log (MSE)) for the complete set of 26 stacks are represented as heatmaps in Figure 7 (a) and (b) for NC and C sub-images, respectively. As the magnitude of noise increases, it becomes harder for the DCNNs to recover features, leading to larger errors between the predicted and measured polarization values. This behavior is somewhat expected. It is also safe to say that for sub-images created with smaller window sizes with higher noises added, the corresponding DCNNs will have much lower performance as compared to that for moderate window sizes. For example, the difference between real and predicted polarizations of sub-images for a window size of 10 and GN = 1000 is much higher as compared to those for a window size of 80 and GN = 1000, as evident from both (a) and (b) heatmaps.

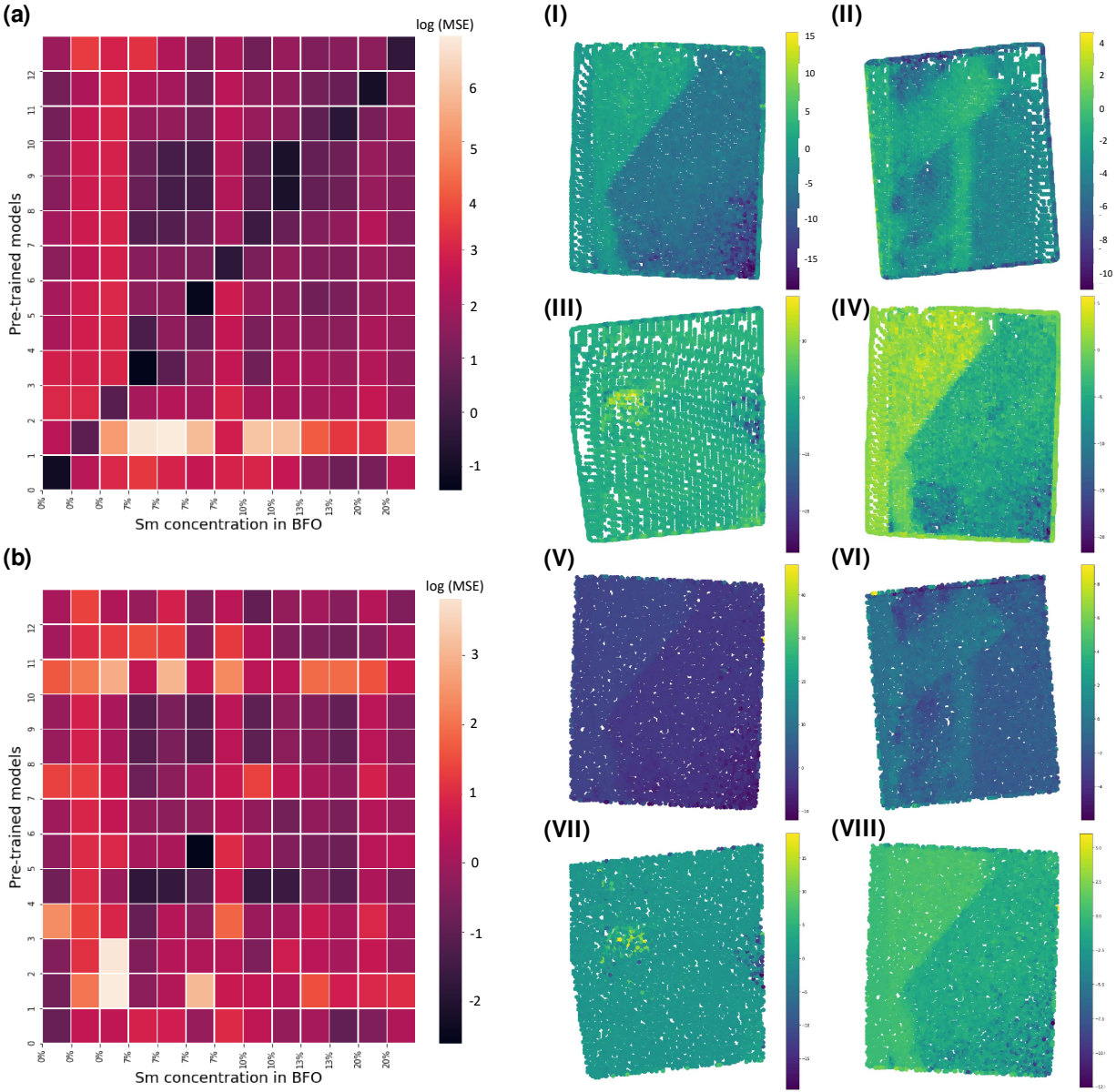

**Figure 8.** Error matrices (a) and (b) generated by plotting MSE values as each of the 13 networks are applied to every 13 sub-image stacks are represented by heatmaps. Predicted polarization maps are displayed in (I-VIII) as DCNN trained on 0% Sm is applied to 0% (I, V) and 7% (II, VI) Sm sub-image stacks as well as DCNN trained on 20% Sm applied to 20% (III, VII) and 0% (IV, VIII) Sm concentrations. While (a) and (I-IV) represent the errors and predicted results for NC, respectively, (b) and (V-VIII) are the same for C.

To evaluate the performance of the trained networks, we computed the MSE for all 26 networks as applied to all 13 sub-image stacks for both NC and C as represented by heatmaps, as shown in Figure 8 (a) and (b), respectively. The stack of sub-images with lesser concentrations of Sm have high polarization values, meaning these systems are more ferroelectric in nature as compared to counterparts with higher dopants concentrations. Therefore, it is expected that deep NNs trained on 0% Sm should exhibit higher performance as applied to system with 20% Sm concentration. However, a network with information on less-ferroelectric to non-ferroelectric systems should fail to predict higher polarization values for the systems with less Sm concentrations. This behavior is further illustrated by Figure 8 (I-IV) and (V-VIII) as a couple of DCNNs trained on 0% and 20% Sm concentrations are applied to (0%, 7%) and (20%, 0%) dopant concentrations, respectively. The diverging colors (lighter to deeper shades) represent low to high polarization values. Figure 8 (I, V) and (II, VI) show how the network trained on ferroelectric image is successful in predicting a range of high-low polarization values. To the contrary, a DCNN trained on sub-images with a 20% Sm concentration only yields reasonable performances when applied to the training set (III, VII) and fails to predict high polarization values.

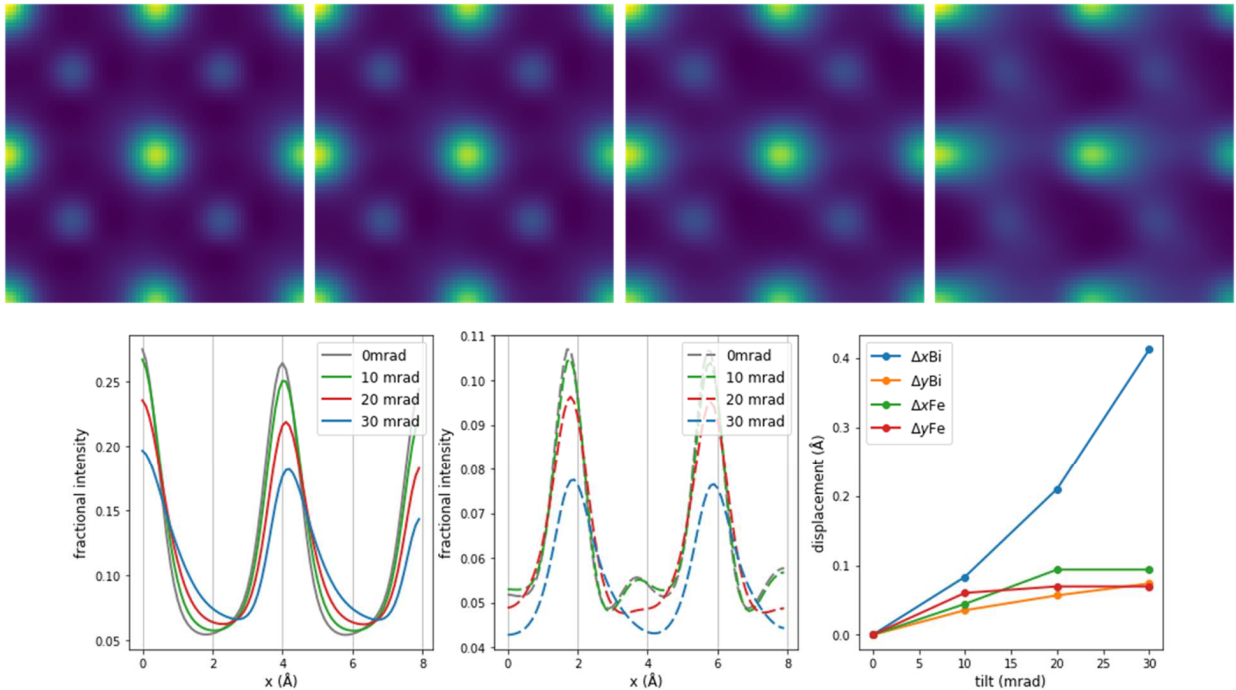

**Figure 9.** Simulated ADF-STEM images of BFO for (a) no tilt, (b) 10 mrad tilt, (c) 20 mrad tilt, and (d) 30 mrad tilt. All tilts are around the y axis. Solid line on (a) represents position of line

scans through Bi columns shown in (e). Dashed line corresponds to position of line scans through Fe columns. Shifts of two atoms circled are shown relative to untilted image shown in (g) for both x and y shift measured using atom finding.

To gain insight into the possible origins of the observed behaviors, including asymmetric polarization distributions and cross-training, we simulated ADF-STEM images for several tilt values off from the zone axis. Calculations were carried out using the $\mu$STEM program and the quantum excitation of phonons algorithm[58]. An accelerating voltage of 200 kV and probe forming aperture of 30 mrad was used. The specimen was assumed to be 100 nm thick and the ADF detector spanned 65—250 mrad. In order to simplify the calculation, the unit-cell angles were adjusted from 59.34 to 60 degrees so it could be converted into a cubic structure that would be more amenable to multi-slice calculations. This small change will have a minimal effect on the qualitative examination of specimen tilt. In Figure 9 (a) the simulated ADF-STEM image for an untilted BFO specimen is shown. Figure 9 (b-d) show increasing tilts with 10 mrad increments; all tilts are clockwise about the vertical y axis. While increasing tilt leads to a reduction in the image contrast, it is unclear to the naked eye if there is a relative shift in the cation positions. To examine this effect, line scans acquired across the Bi columns are shown in Figure 9 (e). For the 30 mrad tilt, a shift of the apparent Bi column is evident. In Figure 9 (f), line scans acquired across the Fe columns are shown and any shifts in the apparent position are much smaller. To quantify this effect, atom finding routines are used to locate the apparent position of the two atoms circled in Figure 9 (a). In Figure 9 (g) we plot the shifts in the x and y directions for each atom relative to the untilted simulation. It is clear that the Bi column shifts approximately twice as far as the Fe column, which would result in an apparent change of polarization.

To summarize, we developed an approach for the analysis of atomically resolved STEM image data of ferroelectric materials to extract local polarization based on sub-image analysis. We demonstrate that the application of DCNN-based regression on sub-images centered on a given sub-lattice yields values similar to direct column position analysis. It should be noted that in both cases, the derived values are biased compared to the expected values. We attribute this behavior to the effect of sample mis-tilt during imaging. Correspondingly, dynamic correction of this effect becomes a key element of the quantitative STEM image and may necessitate the development of tools adjusting the specimen tilt at different parts of the image.

We further show that the polarization fields can be visualized from the STEM images without atom finding using DCNN analysis of atom-centered sub-images and arbitrarily selected sub-images bypassing the atom finding stage. This approach was found to give the correct polarization values for the majority of the image and can be readily incorporated during data acquisition. However, the presence of local defects (i.e., out of distribution data) leads to significant errors in the prediction at certain locations. These can be further used to identify sites for automated experiments. Overall, the translational invariance built in into the DCNN structure can significantly facilitate the extraction of physical order parameter fields from structural and potentially high-dimensional data.

**Acknowledgments**:

This effort (STEM) is based upon work supported by the U.S. Department of Energy (DOE), Office of Science, Basic Energy Sciences (BES), Materials Sciences and Engineering Division (S.V.K., C.T.N.) and was performed and partially supported (M.Z.) at Oak Ridge National Laboratory's Center for Nanophase Materials Sciences (CNMS), a U.S. DOE, Office of Science User Facility. This effort (ML) is based upon work supported by the U.S. DOE, Office of Science, Office of Basic Energy Sciences Data, Artificial Intelligence and Machine Learning at DOE Scientific User Facilities (A.G.). The work at the University of Maryland was supported in part by the National Institute of Standards and Technology Cooperative Agreement 70NANB17H301 and the Center for Spintronic Materials in Advanced Information Technologies (SMART) one of the centers in nCORE, a Semiconductor Research Corporation (SRC) program sponsored by NSF and NIST. The authors gratefully acknowledge Dr. Karren More (CNMS) for careful reading and editing the manuscript.

**Data Availability**:

All the deep learning routines were implemented using a home-built open-source software package AtomAI (https://github.com/pycroscopy/atomai). The dataset is freely available at DOI 10.5281/zenodo.4555978 .
All details of the developed framework are available via two interactive Jupyter notebooks accessible at https://github.com/aghosh92/DCNN_Ferroics .

**Supplementary Materials**

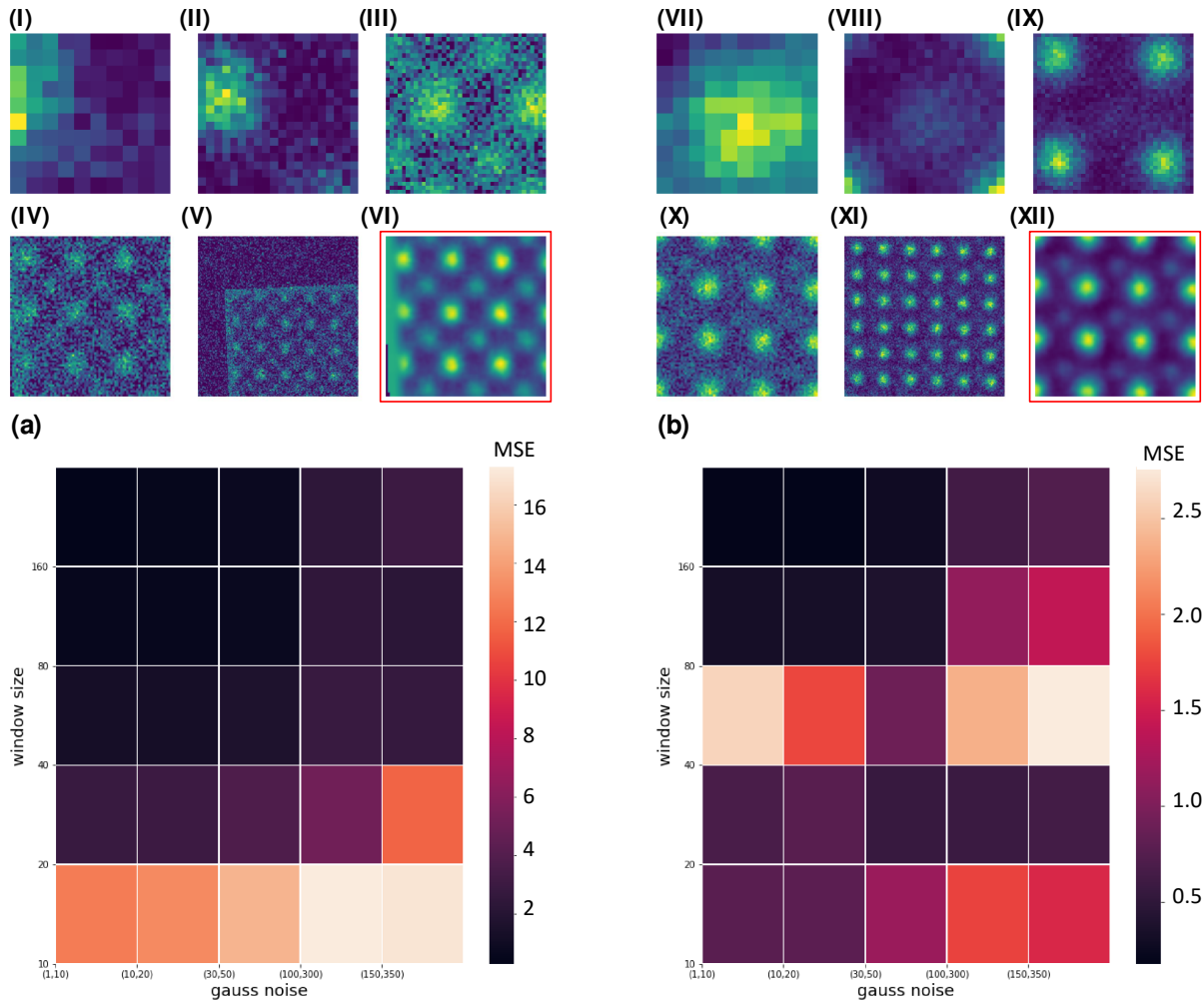

**Figure 1.** Noise sensitivity/performance of DCNNs is shown here for different window sizes (10, 20, 40, 80 and 160) when various levels of gauss noise ((1, 10), (10, 20), (30, 50), (100, 300), (150, 350)) are added to the training dataset. Here each pair represents a range of numbers between which random numbers are picked for each sub-image and corresponding variance is added as gauss noise to the sub-image. Selective sub-images for the series of window sizes and gauss-noises are shown in (I-V) and (VII-XI) for NC and Cs. Reference sub-images are represented by VI and XII corresponding to sub-images generated with no-added noise with window size of 80. The MSE values for each window size-noise combinations are plotted as heatmaps in (a) and (b) for one image for both NC and C, respectively.

To assess the noise and window size sensitivity of DCNNs, a series of networks are trained for five different window sizes (WS) and gauss noises (GN). Selective NC and C sub-images generated with five combinations of WS and GN, such as ([WS- 10, GN – (1,10)], [WS- 20, GN – (10,20)], [WS- 40, GN – (30,50)], [WS- 80, GN – (100,300)], [WS- 160, GN – (150,350)]) are shown in Figure 5 (I-V) and (VII-XI), respectively. Randomized gauss noise between this series of ranges is added to each stack of NC and C subimages. A couple of subimages created using WS of 80 with no noise added are also represented by VI and XII. These are the examples of reference subimages that are utilized for all other analyses in this work. The MSE values for predicted polarization with respect to the observed one for each of the 25 pairs of WS and GN are also computed and plotted as heatmaps in (a) and (b) for NC and C, correspondingly. The test set utilized for these predictions are stack of subimages with respective WS with no noise added. While the analysis shows how a small WS in the range of 10-40 fail to capture the a and b sublattices of the systems, utilizing too big of a window size such as 160 is also not reasonable to capture the atomic features and respective polarization behaviors. Adding any magnitude of noise does not improve the performance of the networks either. In some cases, large amount of added noise makes it difficult for the networks to learn distinct patterns as present in the dataset, that are even obvious to human eyes. We note that for a large enough WS like 80 or 160, the effect of adding more noises seem to be lesser compared to that for smaller WS.